\documentclass[galaxies,article,accept,moreauthors,pdftex]{mdpi} 

\firstpage{1} 
\pubvolume{1}
\issuenum{1}
\articlenumber{0}
\pubyear{2021}
\copyrightyear{2021}
\externaleditor{Academic Editor: {Alexei V. Moiseev} 
} 
\datereceived{} 
\dateaccepted{} 
\datepublished{} 
\hreflink{https://doi.org/} 
\Title{Modelling the Energy Spectra of Radio Relics}

\TitleCitation{Modelling the Energy Spectra of Radio Relics}

\definecolor{myred}{rgb}{0,0.5,0}

\Author{Denis Wittor
 $^{1,2}$*, Matthias Hoeft $^{3}$ and Marcus Br\"uggen $^{1}$}

\AuthorNames{Denis Wittor, Matthias Hoeft and Marcus Br\"uggen}

\AuthorCitation{Wittor, D.; Hoeft, M.; Br\"uggen, M.}

\address{%
$^{1}$ \quad Hamburger Sternwarte, Gojenbergsweg 112, 21029 Hamburg, Germany; mbrueggen@hs.uni-hamburg.de\\
$^{2}$ \quad Dipartimento di Fisica e Astronomia, Universita di Bologna, Via Gobetti 93/2, 40122, Bologna, Italy\\
$^{3}$ \quad Thüringer Landessternwarte, Sternwarte 5, 07778 Tautenburg, Germany; hoeft@tls-tautenburg.de}
\corres{Correspondence: dwittor@hs.uni-hamburg.de}

\abstract{Radio relics are diffuse synchrotron sources 
	 that illuminate shock waves in the in\-tra\-cluster medium. In recent years, radio telescopes have provided detailed observations about relics. Consequently, cosmological simulations of radio relics need to provide a similar amount of detail. 
In this methodological work, we include information on adiabatic compression and expansion, which have been neglected in the past in the modelling of relics.
In a cosmological simulation of a merging galaxy cluster, we follow the energy spectra of shock accelerated cosmic-ray electrons using Lagrangian tracer particles. On board of each tracer particle, we compute the temporal evolution of the energy spectrum under the influence of synchrotron radiation, inverse Compton scattering, and adiabatic compression and expansion. Exploratory tests show that the total radio power and, hence, the integrated radio spectrum are not sensitive to the adiabatic processes. This is attributed to small changes in the compression ratio over time. }


\newcommand{\apjs}{ApJS}

\newcommand{\mnras}{Mon. Not. R. Astron. Soc.}
\newcommand{\aap}{Astron. Astrophys.}

\newcommand{\ssr}{Sci. Space Rev.}

\newcommand{\enzo}{{\it {\small ENZO}}}
\newcommand{\CRaTer}{{\it {\small CRaTer}}}

\newcommand{\music}{{\it {\small MUSIC}}}
\newcommand{\dd}{\mathrm{d}}

\newcommand{\Msun}{\mathrm{M}_{\odot}}

\newcommand{\kpc}{\mathrm{kpc}}

\newcommand{\cm}{\mathrm{cm}}

\newcommand{\sek}{\mathrm{s}}

\newcommand{\Myr}{\mathrm{Myr}}

\newcommand{\erg}{\mathrm{erg}}
\newcommand{\Hz}{\mathrm{Hz}}

\newcommand{\GHz}{\mathrm{GHz}}

\newcommand{\nuobs}{\nu_{\mathrm{obs}}}

\newcommand{\Csync}{C_{\mathrm{sync}}}
\newcommand{\Cadia}{C_{\mathrm{adia}}}
\newcommand{\Cspec}{C_{\mathrm{spec}}}
\newcommand{\Ccool}{C_{\mathrm{cool}}}

\newcommand{\me}{m_\mathrm{e}}
\newcommand{\mpr}{m_\mathrm{p}}
\newcommand{\gram}{\mathrm{g}}
\newcommand{\relicA}{\textit{relic A}}
\newcommand{\relicS}{\textit{relic S}}

\begin{document}

\section{Introduction}
 Radio relics are diffuse radio sources observed at the periphery of galaxy clusters. Relics are the manifestations of cosmic-ray electrons being "shock accelerated" to high energies~\cite{2014IJMPD..2330007B,vanweeren2019review,bykov2019review}. The shock waves develop naturally in the intracluster medium (ICM) during cluster mergers and the accretion of matter. These shock waves (re-) accelerate the ICM electrons to high energies via diffusive shock acceleration (DSA) \cite{1983RPPh...46..973D,ensslin1998}.
 
 As of writing this paper, about 60 relics and candidates are known, and~radio observations are detecting more relics. Furthermore, high-resolution observations show that relics are non-uniform structures and they show threads and filaments on $\kpc$-scales~\cite{rajpurohit2021macsspec,rajpurohit2021A2744}. The~origin of these filaments is still unknown. They might trace magnetic filaments, the~underlying gas density, or the Mach number distribution of the shock~wave.

 {{When computing the cosmic-ray energy spectra in cosmological simulations, previous studies have included energy losses due to synchrotron radiation and inverse Compton scattering only~\cite{2007MNRAS.375...77H,2012MNRAS.420.2006N,2017MNRAS.470..240N,wittor2019pol,wittor2021mach}.}} These works neglected any modification of the energy spectrum due to adiabatic expansion and compression, as done, for example in~\cite{paola1, paola2}. However, if~adiabatic processes are relevant, they might be important when studying the filamentary morphologies of relics. Yet, it is still uncertain if cosmic-ray electrons are subjected to significant adiabatic expansion or compression after they are shock accelerated. It might also be that the post-shock gas dilutes gently. In~this case, the~cosmic-ray electrons are not exposed to any significant adiabatic~processes.
  
 In this contribution, we extend the model derived by~\cite{2007MNRAS.375...77H} and used in~\cite{2012MNRAS.420.2006N,2017MNRAS.470..240N,wittor2019pol,wittor2021mach} to model radio relics. To~this end, we include the effects of adiabatic compression and expansion when computing the cosmic-ray energy spectra. Furthermore, we use Lagrangian tracer particles to follow the spectral evolution of shock accelerated cosmic-ray electrons. We note that this is a methodological paper and we focus on the presentation of the model. An~analysis of the filamentary structures in relics is subjected to future~works.
 
 This paper is structured as follows. In~Section~\ref{2}, we present the cosmic-ray model. In~addition, we present the cosmological simulation that we used. In~Section~\ref{3}, we analyse how adiabatic effects change the total radio power and integrated radio spectra. In~\mbox{Section~\ref{4}}, we summarise and conclude our~work.

\section{Materials and~Methods}\label{2}

In this section, we derive the energy spectrum of shocked accelerated cosmic-ray electrons under the influence of synchrotron radiation, inverse Compton scattering, and~adiabatic compression and expansion.  Furthermore, we present the simulation used in this~work.

\subsection{Modelling of the Cosmic-Ray~Spectra}\label{ssec::spectra}

In~\cite{wittor2019pol,wittor2021mach}, we followed the approach of~\cite{2007MNRAS.375...77H} to model the radio emission of radio relics. Here, we expand our approach by two components. First, the~original model only includes synchrotron and inverse Compton energy losses. Here, we also include adiabatic gains and losses connected to changes of the volume~\cite{1960AZh....37..256S}. Second, in~\cite{wittor2019pol,wittor2021mach}, we computed the electron spectrum using the physical properties at the shock front. This assumption is crude as the physical properties might change in the shock downstream. Hence, we follow the energy spectra using Lagrangian tracer particles, see Section~\ref{ssec::simus}. Consequently, the~new approach is sensitive to local variations of the physical properties as well as adiabatic compression and~expansion.

It is commonly assumed that radio relics are produced by DSA~\cite{ensslin1998}. The~energy spectrum produced by DSA is a power-law distribution.
\begin{align}
  f_0(E) &= \bar{f} E^{-\alpha}. \label{eq::f0}
\end{align}

The energy spectral index depends only on the compression ratio $r$ of the shock and, hence, it is directly connected to the Mach number $M$:
\begin{align}
 \alpha =\frac{r+2}{r-1} = 2\frac{M^2+1}{M^2-1}.
\end{align}

After injection, the~energy spectrum is modified by adiabatic processes as well as synchrotron radiation and inverse Compton scattering. The~modifications are described by~\citep[][]{Kardashev1962}:
\begin{align}
 - \frac{\dd E}{\dd t} = \Csync E^2 + \Cadia E  \label{eq::dedt}.
\end{align}

For consistency with~\cite{2007MNRAS.375...77H}, the~energy is given in units of $\me c^2$: $E\rightarrow E/(\me c^2)$. The~constant $\Csync$ accounts for synchrotron and inverse Compton losses. $\Cadia$ describes adiabatic compression and expansion of the volume $V$:
\begin{align}
 \Csync &= \frac{\sigma_{\mathrm{T}}}{6 \pi m_{\mathrm{e}} c} \left( B_d^2 + B_{\mathrm{CMB}}^2 \right) \\
 \Cadia &= \frac{1}{3 V} \frac{\dd V}{\dd t}
\end{align}

The different variables are: $\sigma_{\mathrm{T}}$ is the Thomson cross section; $\me$ is the electron mass; $c$ is the speed of light; $B_d$ is the downstream magnetic field and $B_{\mathrm{CMB}}$ is the equivalent magnetic field of the cosmic microwave background. The~adiabatic index is $\gamma = 4/3$. In~principle, changes of the cosmic-ray energy due to adiabatic processes can also lead to radio emission, e.g.,~\citep{1981heaa.book.....L}. However, this is expected to be negligible compared to the synchrotron~emission.

To compute the aged energy spectrum, we follow the method described in~\cite{ensslin2001}. Therefore, we introduce the compression ratio
\begin{align}
 \kappa(t) = \frac{V_0}{V(t)} \label{eq::compratio}
\end{align}

which is the ratio of the initial volume $V_0$ and the compressed/expanded volume at time $t$. The~substitution of $e = \kappa^{-1/3} E$ reduces Equation~(\ref{eq::dedt}) to
\begin{align}
 -\kappa(t)^{1/3} \frac{\dd e}{\dd t} = \Csync \kappa(t)^{2/3} e^2.
\end{align}

This equation integrates to
\begin{align}
 e = \frac{1}{\int_{t_0}^{t} \Csync \kappa(t^{\prime})^{1/3} \dd t^{\prime} + k}.
\end{align}

The integration constant is given by $k$. Substituting $E$ back and using the initial condition $E(t_0) = E_0$ gives
\begin{align}
 E = \frac{E_0\kappa(t)^{1/3}}{1 + E_0  \Ccool(t)}.
\end{align}

Here, $\Ccool(t)$ is the integral $\int_{t_0}^t \Csync \kappa(t^{\prime})^{1/3} \dd t^{\prime}$. The~corresponding energy spectrum is computed as
\begin{align}
 f(E,t) = f_0(E_0) \frac{\partial E_0}{\partial E} \kappa(t)
\end{align}

with the initial spectrum Equation~(\ref{eq::f0}) and
\begin{align}
 E_0 &= \frac{E \kappa(t)^{-1/3}}{1- \Ccool(t) \kappa(t)^{-1/3} E}
\end{align}

the energy spectrum becomes
\begin{align}
 f(E,t) = \bar{f} E^{-\alpha} \kappa(t)^{(\alpha+2)/3} \left(1 - \left(\frac{1}{E_{\max}} + \Ccool(t)\kappa(t)^{-1/3}\right) E \right)^{\alpha-2}. \label{eq::spec}
\end{align}

Similar to~\cite{2007MNRAS.375...77H}, we included a high energy cut-off in Equation~(\ref{eq::spec}). The~spectrum is only computed when $\Ccool(t)\kappa(t)^{-1/3}E < 1 - E/E_{\max}$. 
Furthermore, if~adiabatic compression and expansion are neglected, i.e.,~setting the compression ratio $\kappa(t)$ to one, the~spectrum becomes the solution derived by~\cite{2007MNRAS.375...77H}. In~this case, only synchrotron and inverse Compton losses determine the spectrum's shape.
Finally, to~have consistency with~\cite{2007MNRAS.375...77H}, we compute the normalisation of the spectrum as
\begin{align}
 \bar{f} = \frac{n_{\mathrm{e}} \Cspec}{\me c^2} \label{eq::fbar}
\end{align}
with
\begin{align}
 \Cspec &= \xi_{\mathrm{e}} \frac{u_{_\mathrm{d}} \mpr}{c^2 \me}\frac{q-1}{q} \frac{1}{I_{\mathrm{spec}}} \\
 I_{\mathrm{spec}} &= \int\limits_{E_{\min}}^{\infty} E \left(\frac{E}{\me c^2}\right)^{-\alpha} \left(1-\frac{E}{E_{\max}} \right)^{\alpha-2} \dd E .
\end{align}

In the three equations above, the~different variables are: 
$\xi_{\mathrm{e}}$ is the fraction of energy that goes into the acceleration of electrons to suprathermal energies;
$u_{_\mathrm{d}}$ is the internal energy of the downstream gas;
$\mpr$ and $\me$ are the proton and electron masses, respectively;
$c$ is the speed of light;
$q$ is the entropy jump;
$E_{\min} = 10 \mathrm{k_B} T$ is the minimum energy for particles to be considered suprathermal; and $E_{\max} = 10^{10} E_{\min}$ is the maximum energy to which particles are~accelerated.

The similarity with the spectrum derived in~\cite{2007MNRAS.375...77H} allows us to use the same substitution of the variable $\tau$ to compute the corresponding radio emission as the integral of the spectrum and the modified Bessel function:
\begin{align}
 \frac{\dd P}{\dd V \dd \nu} = C_\mathrm{R} \int n(\tau,t) F(1/\tau^2) \dd \tau \label{eq::dPdVdv}
\end{align}
where $\tau$ is a function of $E$ and takes the form:
\begin{align}
 \tau = \sqrt{\frac{3 e B}{16 \nuobs \me c}} \left(\frac{E}{\me c^2} + 1 \right).
\end{align}
\begin{align}
 C_\mathrm{R} = \frac{9 e^{5/2} B^{3/2} \sin(\psi)}{4 \sqrt{\nuobs \me c}}
\end{align}

For the pitch-angle $\psi$, we assume $\sin (\psi) = \pi / 4$.

\subsection{Cluster~Simulation}\label{ssec::simus}

In this work, we combine cosmological simulations carried out with the \enzo \ code~\cite{ENZO_2014} and a Lagrangian analysis performed with the code \CRaTer \ \cite{2017MNRAS.464.4448W}. For~details on the algorithms and implementation, we point to the given~references.

Here, we analyse one cluster from the \textit{SanPedro}-cluster sample. The~\textit{SanPedro}-cluster sample is an on-going campaign to enlarge the number of simulated radio relics. The~final goal is to produce a collection of high-resolution relics that can be compared with observations, e.g.,~\citep{Banfi2020,wittor2020gammas,wittor2021mach}. The~simulation covers a root-grid volume of (140 Mpc/h)$^3$ and is sampled with $256^3$ cells. Five levels of nested refinements and one additional level of adaptive mesh refinements were used to achieve a high and uniform resolution in the (4.85 Mpc)$^3$ volume centred on the cluster. The~resolution is $\Delta x \approx 8.54 \ \kpc/$h on the most refined level. A~uniform magnetic field with $10^{-7} \ \mathrm{G}$ was initialised in each~direction.

\begin{figure}[H]
 \includegraphics[width = 0.74\textwidth]{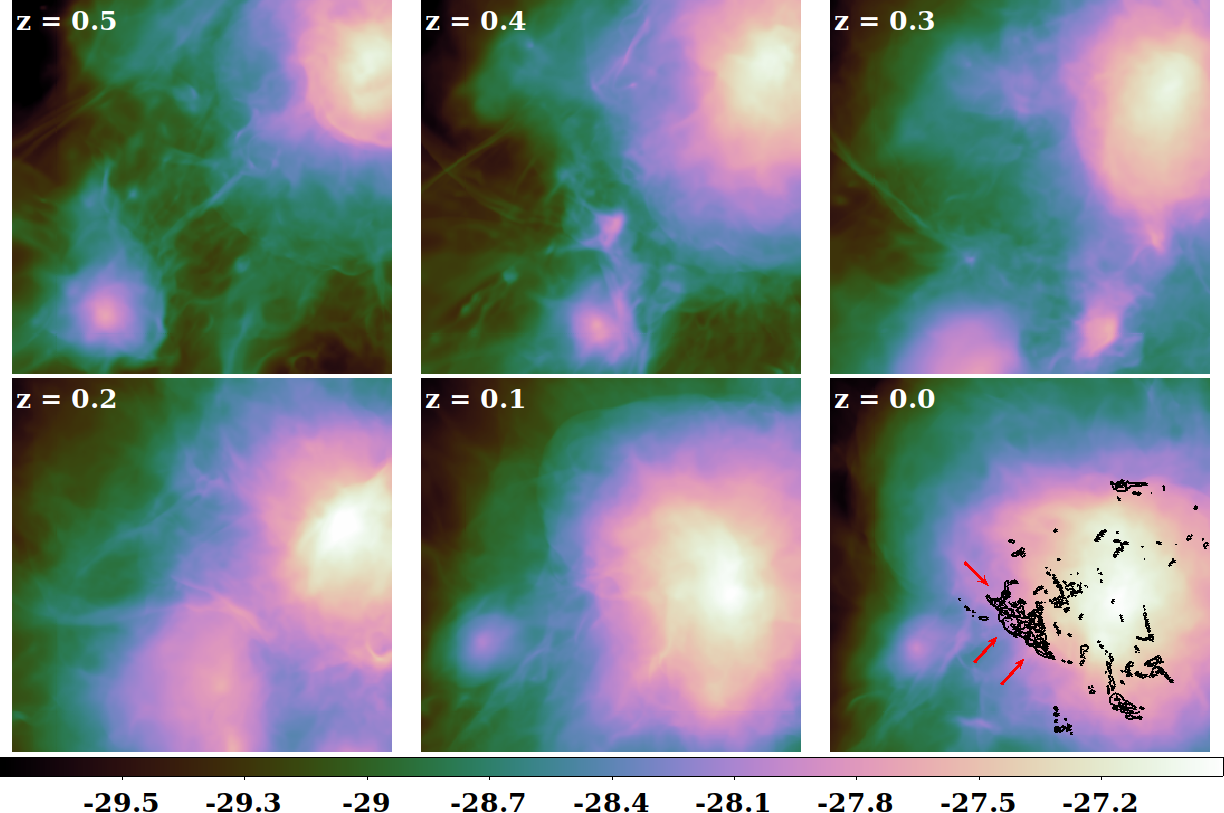}
 \caption{Time-sequence of the cluster merger. The~maps show the projected gas density in units of $\log_{10}(\gram/\cm^3)$ at six different snapshots of the simulation. In~the last panel, the~black contours give the radio power at $[10^{25}, \ 10^{26}, \ \dots, \ 10^{29}] \ \erg/\sek/\Hz$. The~red arrows mark the relic that we study in this work.}
 \label{fig::dens_maps}
\end{figure}

{{The cluster, analysed in this work, is of particular interest as it undergoes a major merger that produces several shock waves. A~large shock wave is developed at redshift $z \approx 0$. The~shock wave is produced by a Mach number distribution that has an average of $\sim$2.8 and a standard deviation of $\sim$1.4. This is a typical shock wave produced during a cluster merger, e.g., compared with Figure~10 in \citep[][]{wittor2021mach}. The~shock wave produces a bright radio relic that has the typical arc-shaped morphology as most observed relics. Hence, the~cluster merger and the radio relic are typical when compared to other simulated and observed radio relics. An~expansion of the analysis to different cluster mergers, including more extreme ones, will be a task for future works. In~Figure~\ref{fig::dens_maps}, we show the evolution of the cluster's baryonic density.}}

To follow the shock injected cosmic-ray energy spectrum, i.e.,~Equation~(\ref{eq::spec}), we use the Lagrangian tracer code \CRaTer \ \cite{2016Galax...4...71W,2017MNRAS.464.4448W,2017MNRAS.471.3212W,wittor2020gammas}. At~redshift $z \approx 1$, we inject $5.1 \cdot 10^5$ passive tracers onto the \enzo-data. Each tracer carries a mass of $m_t = 9.5\cdot10^5 \ \Msun$ and additional tracers are injected at runtime following the mass inflow of the simulation. At~the end of the simulation, the~cluster is sampled with $9.2 \cdot 10^7$ tracers. The~relic itself is sampled by $\sim$5.8 $\cdot 10^4$ tracers.

Using the \textit{cloud-in-cell}-interpolation and an additional correction term to cure for numerical biases in converging flows see~\cite{myphdthesis}, for an elaborated discussion and testing of the scheme, the~tracers are advected across the \enzo-data. The~tracers use a temperature based shock-finder to detect shock waves in the simulated ICM. The~strength of the corresponding shock waves is computed as:
\begin{align}
 M = \sqrt{\frac{4}{5} \frac{T_{\mathrm{new}}}{T_{\mathrm{old}}} \frac{\rho_{\mathrm{new}}}{\rho_{\mathrm{old}}} + 0.2}.
\end{align}

Once a tracer detects a shock, the~injection spectrum, Equation~(\ref{eq::spec}) with $t = 0$, is computed. At~the subsequent timesteps, the~integral $\Ccool(t)$ and the compression ratio $\kappa(t)$ are computed using the grid quantities, recorded by the tracers. Following the \textit{CFL}-condition, the~tracers read the grid quantities and, hence, compute the aged spectrum about every 2--4 $\ \Myr$. The~compression ratio, Equation~(\ref{eq::compratio}), of~the tracers is computed as
\begin{align}
 \kappa(t) = \frac{\rho(t_1)}{\rho(t_2)}. \label{eq::kappa}
\end{align} 

$\rho$ is the local baryonic density measured from the grid at two adjacent timesteps $t_1$ and $t_2$, with~$t_1 < t_2$.

\begin{figure}[H]
 \includegraphics[width = 0.375\textwidth]{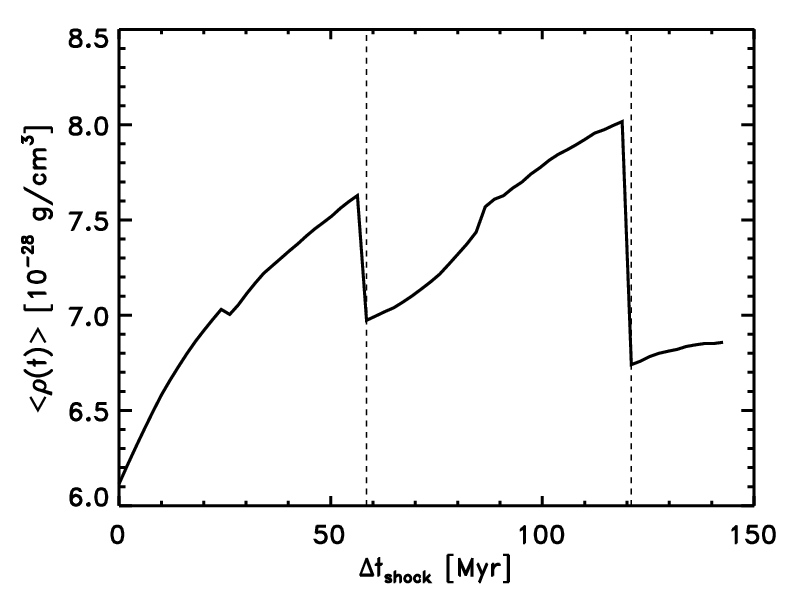}
 \includegraphics[width = 0.375\textwidth]{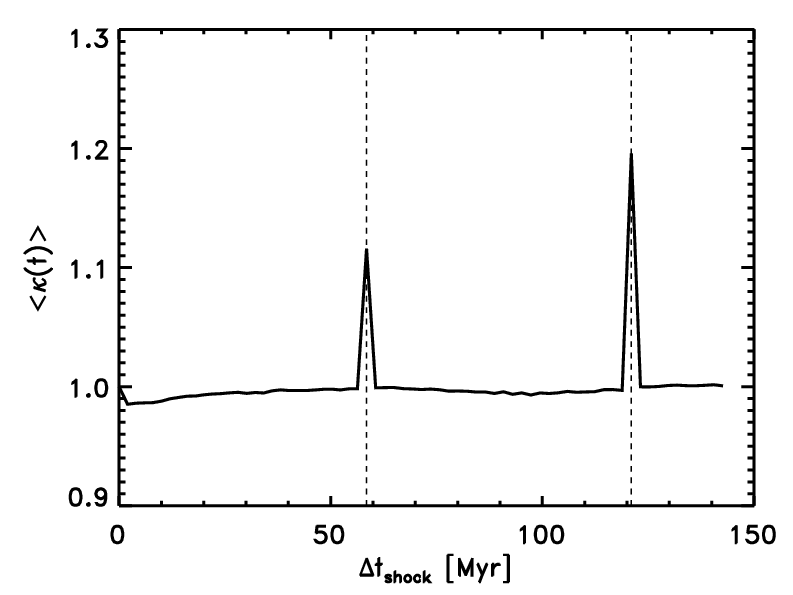}
 \caption{The left panel shows average density measured by the tracers after shock passage, $\Delta t_{\mathrm{shock}} = 0 \ \Myr$. The~right panel shows the average compression ratio measured by the tracers after shock passage. In~both panels, the~two vertical lines mark the timestep, when a new \enzo - snapshot is~loaded.}
 \label{fig::evo}
\end{figure}

In Figure~\ref{fig::evo}, we plot the average density and the average compression ratio measured by the tracers after they have passed the shock at $\Delta t_{\mathrm{shock}} = 0 \ \Myr$. We find that the measured density slowly increases over time, with~two drops at $\sim$60 Myr and $\sim$120 Myr after the shock passage. Consequently, the~average compression ratio is rather small with values around $\sim$1. Only at  $\sim$60 Myr and $\sim$120 Myr, the~comparison ratio increases to $\sim$1.1 and $\sim$1.2, respectively. The~two spikes are of numerical nature, which we explain in the~following.

\CRaTer \ analyses the \enzo-data in post-processing. The~temporal spacing between the saved \enzo-datasets is larger than allowed by the \textit{CFL}-condition. However, for~a stable advection, the~tracers need to follow the \textit{CFL}-condition. Therefore, they are advected over one \enzo-dataset multiple times For an elaborated description of the \CRaTer-implementation, we point the reader to~\cite{myphdthesis}. Once, the~tracers reach the timestep of the next \enzo-dataset, the~new data are loaded and the tracers record the new \enzo-data. Hence, when new \enzo-data are loaded, the~measured tracer data are not only sensitive to the spatial changes due to the tracer advection, but~are also sensitive to temporal changes due to the new \enzo-data. This effect can produce stronger jumps in the tracer data. However, as~seen in Figure~\ref{fig::evo}, this numerical effect is rather~mild.

\unskip

\section{Results}\label{3}

Using the methods described in Section~\ref{ssec::spectra}, we compute the cosmic-ray spectra and the associated radio emission of the shock accelerated particles. Therefore, we distinguish between two models: first, we neglect adiabatic effects and only compute the spectral evolution under synchrotron radiation and inverse Compton scattering. This model is called \relicS. The~second model includes adiabatic effects and is called \relicA. In~the following, we highlight the differences of the two models. Therefore, we focus on the radio power and the integrated radio spectra. Any analysis concerning filamentary structures or the polarisation properties are targets of future~works.

\subsection{Radio~Power}\label{ssec::power}

{{Using Equation~(\ref{eq::dPdVdv}), we compute the radio power at $1.4 \ \GHz$ of the two relics at $\sim 142 \ \Myr$ after the shock passage.}} The radio power of \relicS \ is about $\sim 4.98 \cdot 10^{29} \ \erg/\sek/\Hz$. The~radio power of \relicA \ is about $\sim 4.96 \cdot 10^{29} \ \erg/\sek/\Hz$. Consequently, the~difference of the two is marginal and they can be considered equal. Hence, the~adiabatic effects do not affect the total radio power of the~relic.

\begin{figure}[H]
 \includegraphics[width = 0.75\textwidth]{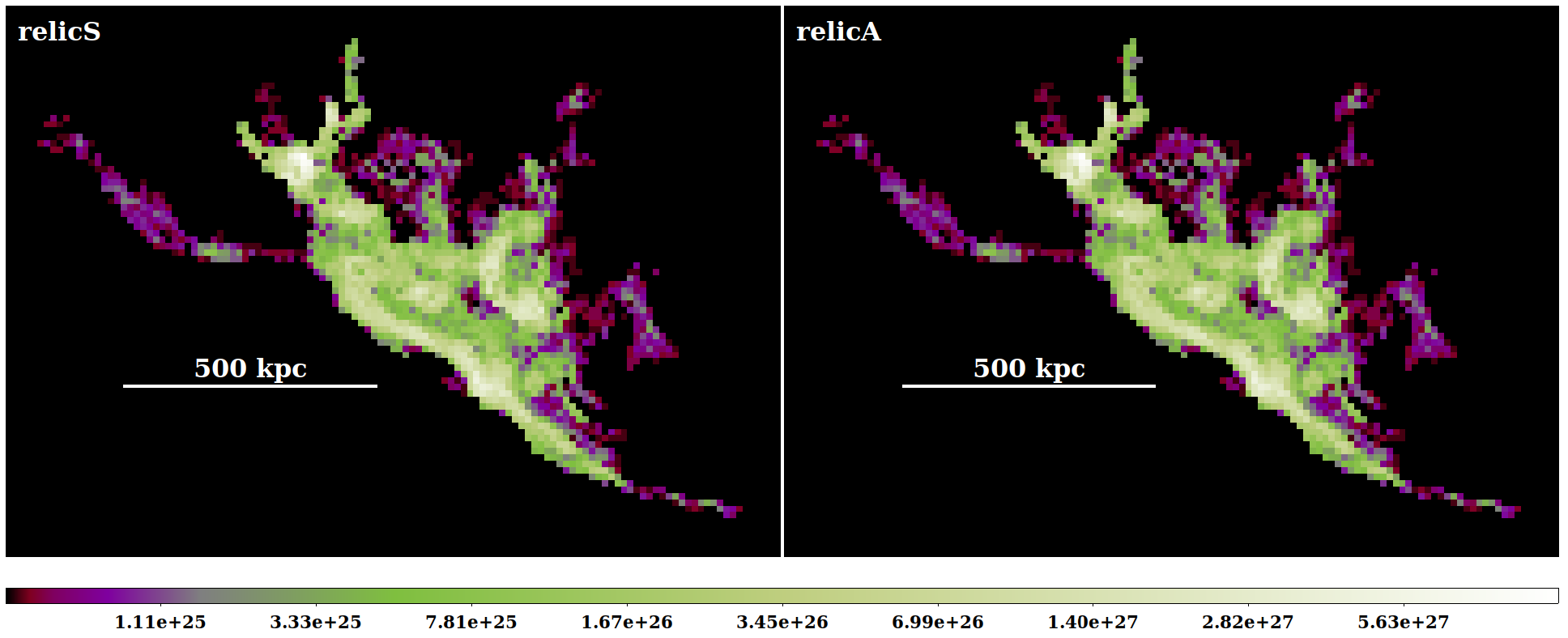}
 \caption{Maps of the radio power for our two models in units of $\erg/\sek/\Hz$. The~left panel shows \relicS. The~right panel shows \relicA.}
 \label{fig::rad_maps}
\end{figure}

Figure~\ref{fig::rad_maps} shows the maps of the radio emission at $1.4 \ \GHz$ of the two models. For~both relics, we find a largest-linear scale of $\sim 1641.80 \ \kpc$. Furthermore, the~relic morphologies appear similar. In~order to investigate the difference of the two maps, we compute a residual map normalised to the map of \relicS, i.e.,~we compute:
\begin{align}
 \Delta P_{1.4 \ \GHz} = \frac{P_{1.4 \ \GHz}(\mathrm{relic \ S})-P_{1.4 \ \GHz}(\mathrm{relic \ A})}{P_{1.4 \ \GHz}(\mathrm{relic \ S})} . \label{eq::res}
\end{align}

\begin{figure}[H]
 \includegraphics[width = 0.375\textwidth]{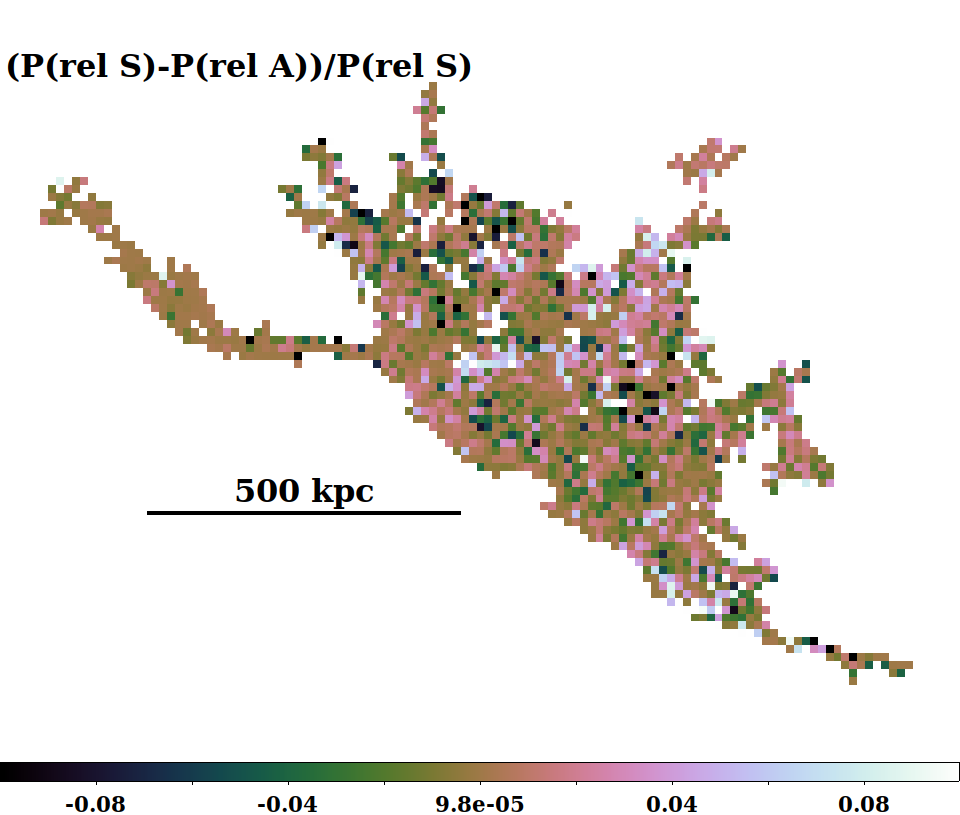}
 \includegraphics[width = 0.375\textwidth]{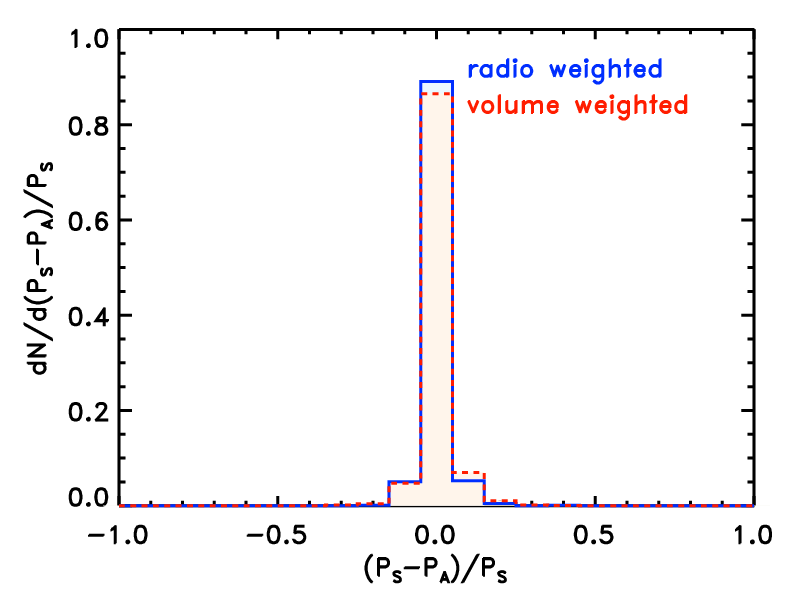}
 \caption{The left panel shows the residual map of the two relics (see Equation~(\ref{eq::res})). The~right panel shows the volume weighted and radio-weighted distributions of the~residuals.}
 \label{fig::diff_maps}
\end{figure}

The residual map is shown in the first panel of Figure~\ref{fig::diff_maps}. For~the most part, the~difference between the two maps is small and approaches zero. Only in a few regions that are mostly located in the far downstream do we find significant differences between the~two. 

To better quantify the difference between the two models, we computed the pixel-weighted and radio-weighted distribution of the residual map. The~distributions are given in the second panel of Figure~\ref{fig::diff_maps}. The~pixel weighted distribution peaks around zero, with~$\sim 85 \ \%$ of the pixels showing a difference between $-0.1$ and $0.1$. For~the radio-weighted distribution, the peak between $-0.1$ and $0.1$ increases above $\gtrsim 90 \ \%$. Consequently, the~relative difference between \relicA \ and \relicS \  is minor in the radio bright regions. Vice~versa, the~relative difference is larger in fainter~regions. 

The results above show that the adiabatic effects are neglectable when studying the total radio power and maps of radio~relics.

\unskip

\unskip

\subsection{Integrated Radio~Spectra}

\begin{figure}[H]
 \includegraphics[width = 0.375\textwidth]{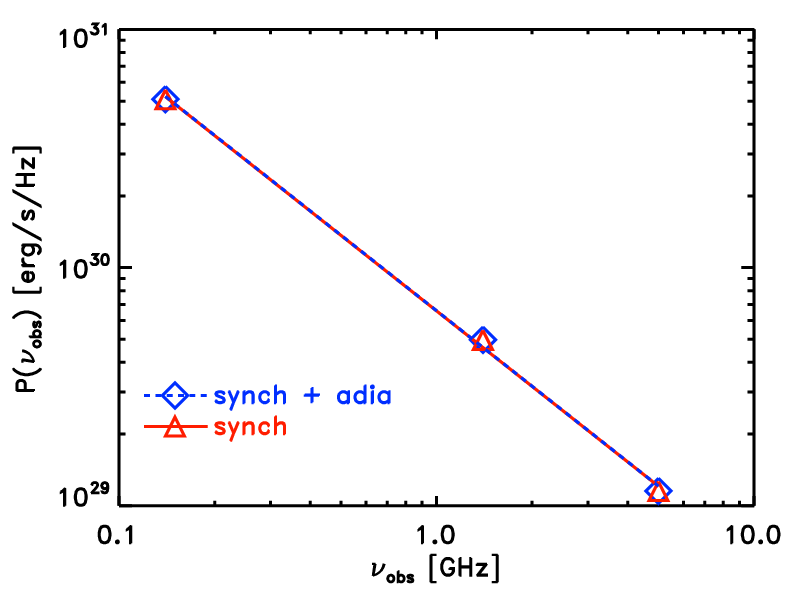}
 \includegraphics[width = 0.375\textwidth]{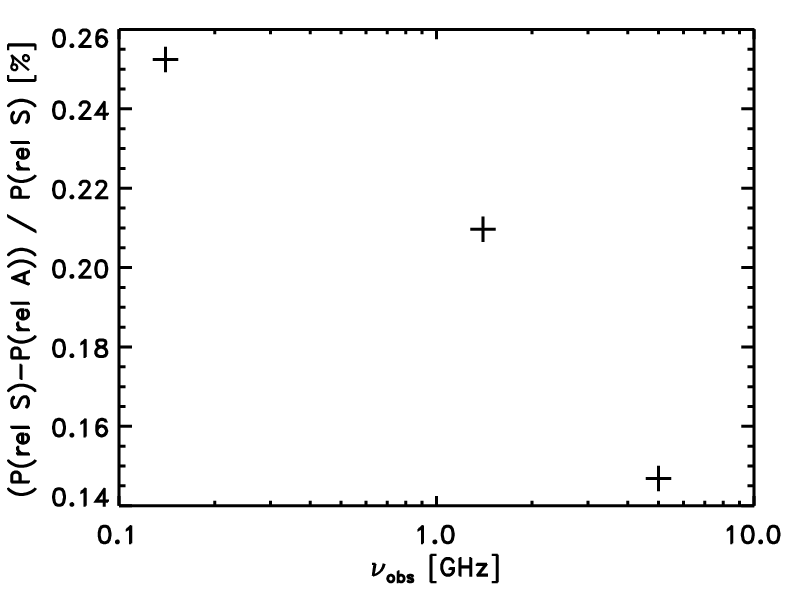}
 \caption{\textls[-15]{Left panel: Comparison of the integrated radio spectra measured at $\nuobs = [0.14,1.4,5.0] \ \GHz$.} The~red triangles give the spectrum of \relicS. The~blue diamonds give the spectrum for \relicA. The~red solid line and the blue dashed line give the fitted power-laws. Right panel: relative difference of the integrated radio~spectra.}
 \label{fig::spec_comp}
\end{figure}

To investigate the effect on observing frequencies other than $1.4 \ \GHz$, we computed the radio emission at $0.14 \ \GHz$ and $5.0 \ \GHz$. Using the radio power at the three different frequencies, we computed the integrated radio spectra for both relics. The~radio spectra are plotted in Figure~\ref{fig::spec_comp}. Here, we use the convention that $I(\nu) \propto \nu^{\alpha}$

Fitting a power-law to the spectra, we find the same spectral index for both \relicA \ and \relicS. The~spectral indices are $\alpha \approx -1.05$. The~fitted spectral indices start to diverge at the fourth decimal place and, hence, they can be considered equal. Consequently, the~integrated radio spectrum remains unaffected by adiabatic~effects.  

In Figure~\ref{fig::spec_comp}, we also show the relative difference of the radio power of the two relics at the three frequencies. We find that independent of the frequency, the~radio power differ by less than $0.26 \ \%$. This explains the similar spectral~indices.

{{The integrated radio spectrum plotted in Figure~\ref{fig::spec_comp}, shows a nearly perfect power-law over a large frequency range. In~principle, relics should show a high frequency cut-off because they are affected by radiation cooling. However, in~the quasi-stationary shock scenario, as~analysed here, a~power-law spectra is expected. Nevertheless, \citet{loi2020sausspec} and \citet{rajpurohit2020toothspec} produced the most detailed integrated radio spectra for the sausage relic and toothbrush relic, respectively. These observations cover an even larger frequency range, i.e., up to $18.6 \ \GHz$. Yet, neither of them observed a high frequency cut-off. \citet{rajpurohit2020toothbrush} argued that the high Mach number tail of the distribution producing the relic dominates the spectral distribution.}}

{{The integrated spectra were computed at at $\sim$142 Myr after the shock passage. During~this period, the~average compression ratio is very mild, $\langle \kappa (142 \ \Myr) \rangle \approx 1$. Despite its numerical nature, we investigated if~the strongest change in density at $\sim 120 \ \Myr$ affected the total radio power and the integrated spectra. During~this period, the~radio power and, hence, the~spectral index differed more. This relative difference between the radio power of \relicA \ and \relicS \ is the largest at low frequencies: at 0.14 GHz, the~relative difference is about $\sim$8.7\%; at 1.4 GHz, it is about $\sim$2.3\%; and at 5 GHz, it is about $\sim$0.3\%. Hence, the~integrated spectral index for \relicA \ and \relicS \ differs as well. The~relative difference of the integrated spectral index was $\sim$2.8\%. Consequently, for a larger compression ratio, the two models did not differ significantly.}}

\unskip

\subsection{Numerical~Caveats}\label{ssec::caveats}

{{Our results show that adiabatic effects are most likely negligible when mo\-del\-ling radio relics. However, these results could also vary with the spatial resolution, the~number of tracers, or the time resolution of the simulation. Hence, we performed a few tests to determine the corresponding effects.}}

{{First, we reconstructed the \enzo-data at a higher AMR level, with~a resolution}} of $\sim$17.08 kpc/h. {{Following \ref{ssec::simus}, we analysed this simulation with $\sim$2.6 $\cdot 10^7$ tracer particles. However, we find that the differences in radio between \relicA \ and \relicS \ are infinitesimal small. This is due to the fact that the spatial resolution decreases, while the temporal resolution remains the same. Hence, the~tracer particles remain in a volume of constant density}}  for a longer time than the tracer particles in the high-resolution simulation. We note that the density remains constant within a simulation cell for one timestep of the \enzo-simulation. {{Hence, a~high spatial resolution is needed, when studying adia\-batic effects. They might become even more relevant at an even higher resolution. Following the results at $120 \ \Myr$, we do not expect adiabatic processes to become significantly relevant at a higher-resolution. Yet, this needs to be explored using tailored simulations. }}

{{Second, we re-simulated the simulation described in Section~\ref{ssec::simus} using twice the mass resolution of tracers, i.e.,~$m_t = 19 \cdot 10^5 \ \Msun$, for~a total of $\sim 1.9 \cdot 10^7$ tracer particles at the end of the simulation. At~1.4 GHz, the~relic has a total power of $\sim 5.95 \cdot 10^{29} \ \erg / \sek / \Hz$ and $\sim 5.93 \cdot 10^{29} \ \erg / \sek / \Hz$ for \relicS \ and \relicA, respectively}} We note that the radio power is higher than in Section~\ref{ssec::power}. We attribute this to the larger normalisation, Equation~(\ref{eq::fbar}). {{Consequently, the~radio power at 1.4 GHz can be considered equal between the two models. Moreover, the integrated spectral index does not differ between the two models. Hence, we conclude that the number of tracer particles does not increase or decrease the relevance of adiabatic processes.}}

{{Finally, we are already using the highest time resolution possible. Hence, we cannot increase it any further. However, we argue that a higher temporal resolution would not alter the results. The~tracer particles record the compression ratio, Equation~(\ref{eq::kappa}), as~the ratio between densities measured at two adjacent timesteps. If~this ratio was measured at a higher temporal resolution, the~change in density would be smaller or the same, but~never larger. Hence, the~measured compression could only be milder or the same. Consequently, the~adiabatic effects on the spectrum will be the same or less.}}

\section{Conclusions}\label{4}

In this work, we included adiabatic expansion and compression when mo\-del\-ling radio relics. Therefore, we followed the approach of~\cite{ensslin2001} to compute the aged electron spectrum, see Equations~(\ref{eq::dedt}) and (\ref{eq::spec}). Following, reference~\cite{2007MNRAS.375...77H} we computed the associated radio emission. The~inclusion of adiabatic processes might be relevant, when studying the filamentary structures observed in several relics. {{In these structures, the~volume differences might be larger and, hence, compression might be more relevant. Moreover, the~filamentary structures evolve on timescales that are shorter than the cooling times. Hence, the~role of adiabatic processes might be relevant on small timescales.}} {{Adiabatic processes are also relevant in the radio phoenices scenario, e.g.,~\citep{ensslin2001,2002MNRAS.331.1011E,2015MNRAS.448.2197D,2020A&A...634A...4M}, which cannot be studied with the current simulation set-up.}} We stress that this is meant to be a methodological work and that we focus on the presentation of the model. Hence, this study lays the groundwork for future studies of the filamentary structures observed in~relics.

To investigate the effects of adiabatic processes, we analysed a simulated galaxy cluster from the \textit{SanPedro}-cluster catalogue. We used the Lagrangian tracer code \CRaTer \ to follow the shock injected electron spectrum and the subsequent modifications of the spectrum due to synchrotron and inverse Compton losses, as~well as adiabatic compression and expansion, see Section~\ref{2}. The~usage of Lagrangian tracer particles allowed us to compute the changes of the energy spectrum based on the local gas properties. The~latter is also an improvement compared to our previous works~\cite{wittor2019pol,wittor2021mach}. {{Here, we analysed a typical cluster merger. The~shock wave that produces the relic was produced by a distribution with an average of $\sim$2.8 and a standard deviation of $\sim$1.4, e.g., compared with Figure~10 in~\citep{wittor2021mach}. In~the future, it will be useful to conduct a parameter study for mergers of different strengths.}}

To this end, we modelled the relic emission twice, see Section~\ref{3}. In~the first model, named \relicS, we neglected adiabatic processes, while in the second model, named \relicA, adiabatic processes were included. To~find differences between the two models, we compared: the total radio power at $1.4 \ \GHz$, maps of the radio power at $1.4 \ \GHz$, and~the integrated radio spectrum measured between $0.14 \ \GHz$, $1.4 \ \GHz$, and~$5.0 \ \GHz$. For~all three proxies, we found marginal differences between the two~models. 

These findings are attributed to the small changes in the compression ratio that mostly oscillates around $\sim$1, see Figure~\ref{fig::evo}. This indicates that the post-shock gas dilutes mildly. Hence, the~adiabatic compression and expansion are rather unimportant when computing the cosmic-ray spectrum, see $\kappa(t)$ in Equation~(\ref{eq::spec}). {{These results might change with a higher spatial resolution, see Section~\ref{ssec::caveats}. This needs to be investigated in the future.}} {{However, at~$\sim$60 Myr and $\sim$120 Myr, the~simulation shows two spikes of stronger compression. Even though, these spikes are due to numerical reasons, they can be used to estimate the effect of events of strong compression. During~the periods, the~radio power of \relicA \ and \relicS \ start to diverge more significantly at low frequencies. On~the other hand, the~radio powers at high frequencies remain similar. Consequently, the~integrated radio spectrum differs between \relicS \ and \relicA. Yet, the~relative difference of the integrated spectral index is small, $\sim$2.7\%.}} Hence, we conclude that adiabatic effects are most likely negligible when modelling radio~relics.

\vspace{6pt} 
\authorcontributions{D.W. ran the simulation, implemented the codes, and performed the data analysis. D.W., M.H., and M.B. analysed and interpreted the~data. All authors have read and agreed to the published version of the manuscript.}

\funding{D.W. is funded by the Deutsche Forschungsgemeinschaft (DFG, German Research Foundation)---441694982. 
M.H. acknowledges support by the BMBF Verbundforschung under the grant 05A20STA.
M.B. acknowledges funding by the Deutsche Forschungsgemeinschaft (DFG, German Research Foundation) under Germany's Excellence Strategy---EXC 2121 ``Quantum Universe''---390833306.}
\institutionalreview{Not applicable}
\informedconsent{Not applicable}

\dataavailability{The data presented in this is study are part of an ongoing study, and are too large to be archived; hence, the~data are not publicly available yet. However, we are happy to share the data upon~request.} 

\acknowledgments{The authors gratefully acknowledge the Gauss Centre for Supercomputing e.V. (\url{www.gauss-centre.eu}) for supporting this project by providing computing time through the John von Neumann Institute for Computing (NIC) on the GCS Supercomputer JUWELS at Jülich Supercomputing Centre (JSC), under~project no.~hhh44.
The cosmological simulations used in this work were initialised and performed with the \music-code \url{https://bitbucket.org/ohahn/music/wiki/Home} \citep{music} and the \enzo-code \url{http://enzo-project.org} \citep{ENZO_2014}, respectively. 
This research made use of the radio astronomical database galaxyclusters.com, maintained by the Observatory of~Hamburg. }

\conflictsofinterest{The authors declare no conflict of interest. The~funders had no role in the design of the study; in the collection, analyses, or interpretation of the data; in the writing of the manuscript, or~in the decision to publish in this~work.}

\end{paracol}

\reftitle{References}

\end{document}